\documentstyle[twoside,epsfig]{adacta}
\begin{document}

\prvastrana=1
\poslednastrana=5

\def\autor{K. Banaszek}
\def\nazov{Reconstruction of photon distribution}

\headings{1}{5}

\title{RECONSTRUCTION OF PHOTON DISTRIBUTION WITH~POSITIVITY CONSTRAINTS}
\author{{Konrad Banaszek}\footnote{\email{Konrad.Banaszek@fuw.edu.pl}}}
{Instytut Fizyki Teoretycznej, Uniwersytet Warszawski, Ho\.{z}a~69,
PL--00--681~Warszawa, Poland}

\datumy{}{}

\abstract{An iterative algorithm for reconstructing the photon
distribution from the random phase homodyne statistics is discussed. This
method, derived from the maximum-likelihood approach, yields a positive
definite estimate for the photon distribution with bounded statistical
errors even if numerical compensation for the detector imperfection
is applied.}

A fascinating topic studied extensively in quantum optics over past
several years is the measurement of the quantum state of simple
physical systems [1]. The central question is how to reconstruct a family
of observables characterizing the quantum state from data that can be
obtained using a feasible experimental scheme [2]. Most of the
initial work on this topic was based on considerations of ideal,
noise-free distributions of quantum observables that can be obtained
only in the limit of an infinite number of measurements. However, any
realistic experiment yields only a finite sample of data. Recognition
of the full importance of this fact has led to the development of
reconstruction techniques specifically designed to deal with finite
ensembles of physical systems [3--6]. The main motivation for these
developments is to optimize the amount of information on the quantum
state that can be gained from a realistic measurement, and to
distinguish clearly between the actual data obtained from an
experiment and {\em a priori} assumptions used in the reconstruction
scheme.

One of ideas that have proved to be fruitful in the field of quantum
state measurement is the principle of maximum-likelihood estimation.
In particular, it has been recently applied to the reconstruction of
the photon distribution from the random phase homodyne statistics [7]. The
essential advantage of the maximum-likelihood technique is that
physical constraints on the quantities to be determined can be
consistently built into the reconstruction scheme. This reduces the
overall statistical error, and automatically suppresses unphysical
artifacts generated by standard linear reconstruction techniques [8],
such as negative occupation probabilities of Fock states.  However,
this is achieved at a significantly higher numerical cost than that
required by linear techniques.

Before we pass to the detailed discussion of the maximum-likelihood
method, let us demonstrate its capability to improve the accuracy of
the reconstruction. Fig.~1 depicts the final result of processing
Monte Carlo simulated homodyne data for a coherent state and a squeezed
vacuum state; both states have the average photon number equal to
one. We have assumed imperfect photodetection characterized by a quantum
efficiency $\eta=85\%$, which is numerically
compensated in the reconstruction scheme.
It is seen that the maximum-likelihood estimate, marked with dots,
is much closer to the exact distribution than probabilities obtained
using the standard linear method of pattern functions. Moreover, purely
artificial nonzero values for large photon numbers completely disappear
when the maximum-likelihood method is applied.

\begin{figure}
\centerline{\epsfig{file=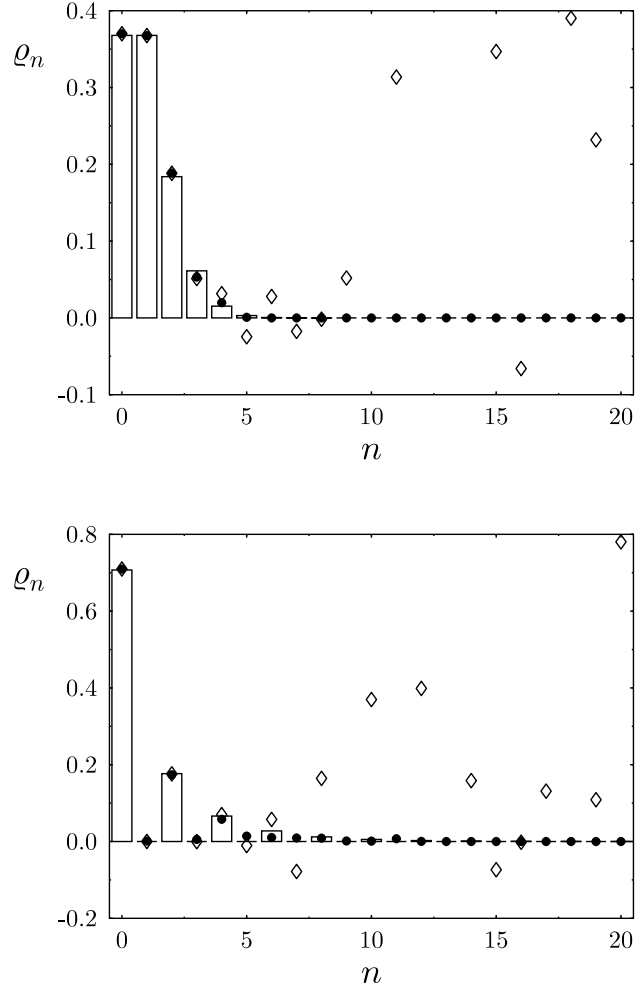,height=14cm}}

\bigskip

\caption{Fig.~1. Reconstruction of the photon distribution from Monte
Carlo homodyne data. The bars depict exact photon distributions for a
coherent state (top) and a squeezed vacuum state (bottom), both the
states with the average photon number equal to one. For each of these
states, $N=10^5$ homodyne events were simulated. The events were
divided into 100 bins covering the interval $-5 \le q \le 5$.  The
simulated data were used to reconstruct the photon distributions using
the maximum-likelihood method ($\bullet$) and the linear pattern
function technique ({\small $\diamondsuit$}). In the latter case, some of
reconstructed $\varrho_n$'s are beyond the scale of the graphs. 
Note that for small $n$, Fock state occupation probabilities reconstructed
using both the methods are very close, and the symbols $\bullet$
and {\small $\diamondsuit$} overlap.  The
maximum-likelihood estimates were obtained from 8000 iterations of the
EM algorithm, starting from a uniform distribution for $0\le n\le 20$.}
\end{figure}

So, how does the maximum-likelihood method work, and what algorithm
provides the bounded, positive definite estimate for the photon
distribution? Let us start the discussion from considering
the raw data recorded in
an experiment. A single experimental run of a random phase homodyne
setup yields a certain value of the quadrature observable $q$. This data,
after an appropriate discretization, is recorded as an event in a $\nu$th
bin. The probability of this event $p_\nu$ depends linearly on the
photon distribution $\{\varrho_n\}$:
\begin{equation} 
p_\nu(\{\varrho_n\}) = \sum_{n} A_{\nu n} \varrho_{n}.
\end{equation} 
For a fixed $n$, the set of coefficients $A_{\nu n}$ describes the
homodyne statistics for the $n$th Fock state.

Repeating the measurement $N$ times yields a frequency histogram $\{k_n\}$
specifying in how many runs the outcome was found in a $\nu$th bin.
These incomplete, finite data are the source of information on the photon
distribution. The question is, how this information can be retrieved.
The answer given by the maximum-likelihood estimation method is to pick up the
photon distribution for which it was the most likely to obtain the
actual result of the performed series of measurements. Mathematically,
this is done by the maximization of the log-likelihood function [7]:
\begin{equation}
{\cal L}(\{\varrho_n\}) = \sum_{\nu} k_{\nu} \ln p_{\nu} (\{\varrho_n\})
- N \sum_{n} \varrho_{n},
\end{equation}
where $N=\sum_{\nu} k_{\nu}$ is the total number of experimental runs.
In the above formula, the method of Lagrange multipliers has been used to
satisfy the condition that the sum of all probabilities is equal to one.
As the estimate for $\{\varrho_n\}$ is supposed to describe a possible
physical situation, the search for the maximum likelihood is {\em a
priori} restricted to the manifold of real probability distributions. 
Geometrically, this manifold is a simplex defined by the set of 
inequalities:
\[ \varrho_{n} \ge 0, \hspace{1cm} n=0,1,2,\ldots 
\]\[
\sum_{n} \varrho_{n} = 1.
\]
Thus, the physical constraints on the reconstructed quantities are 
naturally incorporated in the maximum-likelihood reconstruction scheme. 

The reconstruction of the photon distribution formulated in the
maximum-likelihood approach belongs a very wide class of linear inverse
problems with positivity constraints~[9]. This class encompasses a variety
of problems appearing in as diverse fields as medical imaging and
financial markets. As discussed by Vardi and Lee [9], a
straightforward method for solving these problems is provided by the
so-called expectation-maximization (EM) algorithm. In the following, we
will present a heuristic derivation of this algorithm applied the
reconstruction of the photon distribution.

Let us consider the necessary condition for the maximum of the
function ${\cal L}(\{\varrho_n\})$.
The partial derivatives of the log-likelihood function are given by:
\begin{equation}
\frac{\partial {\cal L}}{\partial\varrho_m} = 
\sum_{\nu} k_{\nu} \frac{A_{\nu m}}{p_\nu(\{\varrho_n\})} - N
\end{equation}
For each $m$, the partial derivative ${\partial {\cal
L}}/{\partial\varrho_m}$ must vanish unless the maximum is located on
a face of the simplex for which $\varrho_m = 0$.  Thus we have that
either ${\partial {\cal L}}/{\partial\varrho_m}=0$ or $\varrho_m = 0$.
These two possibilities can be written jointly as
\begin{equation}
\varrho_m \left(\sum_{\nu} k_{\nu} 
\frac{A_{\nu m}}{p_\nu(\{\varrho_n\})} - N\right)
=0, \hspace{1cm} m=0,1,2,\ldots.
\end{equation}
It is convenient to rearrange this condition to the form which defines 
the maximum-likelihood estimate as a fixed point of a certain nonlinear
transformation. Such a form immediately suggests an iterative procedure
for finding the fixed point by a multiple application of the derived
transformation:
\begin{equation}
\varrho_{m}^{(i+1)} = \varrho_{m}^{(i)} \sum_{\nu}
\frac{k_\nu}{N} \frac{A_{\nu m}}{p_{\nu}(\{\varrho_n^{(i)}\})},
\hspace{1cm} m=0,1,2,\ldots
\end{equation}
where upper indices in parentheses denote consecutive approximations
for the photon distribution.

In fact, Eq.~(5) provides a ready-to-use iterative method for
reconstructing the photon distribution, which is a special case of the
EM algorithm [9]. Given an approximation of the photon distribution
$\{\varrho_n^{(i)}\}$, the next one is simply evaluated according to
the right hand side of Eq.~(5).  When sufficient mathematical
conditions are fulfilled, this procedure converges to the
maximum-likelihood solution.  Thus, the complex problem of constrained
multidimensional optimization is effectively solved with the help
of a simple iterative algorithm.
The maximum-likelihood estimates depicted in Fig.~1 were obtained from
8000 iterations of this algorithm. The initial distributions were
assumed to be uniform for photon numbers $n \le 20$ and equal to
zero above this cut-off value.

\begin{figure}
\centerline{\epsfig{file=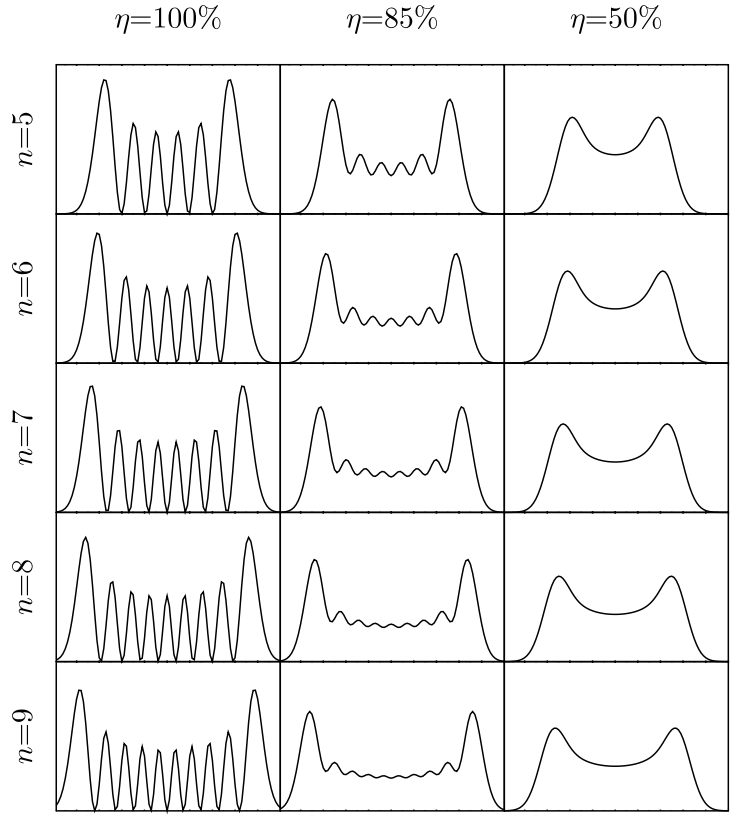,width=100mm}}

\bigskip

\caption{Fig.~2. 
Components of the random phase homodyne statistics generated by different
Fock states, for decreasing efficiency $\eta$ of the homodyne detector.
The homodyne statistics of an arbitrary quantum state is a sum of these 
components with weights defined by the photon distribution $\{\varrho_n\}$.}
\end{figure}

An essential yet delicate matter in quantum state measurements is the
role played by the detection efficiency. The impact of imperfect
detection on the maximum-likelihood reconstruction scheme can be
understood using a simple intuitive argument.  According to Eq.~(1),
the homodyne statistics is a sum of components generated by the Fock
states $|n\rangle$, with the weights given by the appropriate
occupation probabilities $\varrho_n$. The shape of each component is
described by the set of coefficients $A_{\nu n}$, with $n$ treated as a
fixed parameter. Fig.~2 depicts several of these components for
different values of the detector efficiency $\eta$. In the case of unit
efficiency, the contribution originating from the Fock state
$|n\rangle$ is given by the squared modulus of the $n$th energy
eigenfunction in the position representation, and exhibits
characteristic oscillations.  The important point is that each of these
contributions has a unique location of maxima and minima.  Thus, each
number state leaves a specific, easily distinguishable trace in the
homodyne statistics.  Roughly speaking, this is what allows the
maximum-likelihood method to resolve clearly the contributions from
different number states. When the detection is imperfect, homodyne
statistics generated by Fock states become blurred, and oscillations
disappear. This makes the shape of the contributions from higher Fock
states quite similar. Consequently, it becomes more difficult for the
maximum-likelihood method to resolve components generated by different
number states.  Of course, this intuitive argument should be supported
by quantitative considerations.  Mathematically, the effect of imperfect
detection is reflected by the shape of the log-likelihood function
${\cal L}$. For low detection efficiency, it becomes flatter, and its
maximum is not sharply peaked. This increases the statistical
uncertainty of the reconstructed photon distribution, and results in a
slower convergence of the EM algorithm.

Finally, let us note that the maximum-likelihood algorithm does not
make any use of the specific form of the coefficients $A_{\nu n}$
linking the photon distribution with the homodyne statistics. In fact,
it can be applied to any phase-insensitive measurement whose
statistical outcome depends on the photon distribution via a linear
relation of the form assumed in Eq.~(1).

\medskip

\noindent {\bf Acknowledgements}
The author has benefited from discussions with Prof.\ Krzysztof
W\'{o}dkiewicz and Dr.\ Arkadiusz Or{\l}owski. This research was supported
by Komitet Bada\'{n} Naukowych and by Stypendium Krajowe dla M{\l}odych
Naukowc\'{o}w Fundacji na rzecz Nauki Polskiej.

\small
\kapitola{References}
\begin{description}
\itemsep0pt
\item{[1]}
See, for example, {\sl J. Mod. Opt.} {\bf 44} (1997) No.\ 11-12, Special
Issue on Quantum State Preparation and Measurement, ed.\ by M.~G.~Raymer
and W.~P.~Schleich.
\item{[2]}
\refer{G. M. D'Ariano, U. Leonhardt, H. Paul}{Phys. Rev. A}{52}{1995}{R1801}
\item{[3]} 
\refer{V. Bu\v{z}ek, G. Adam, G. Drobn\'{y}}{Phys. Rev. A}{54}{1996}{804}
\item{[4]}
\refer{T. Opatrn\'{y}, D.-G. Welsch, W. Vogel}{Phys. Rev. A}{56}{1997}{1788}
\item{[5]}
\refer{Z. Hradil}{Phys. Rev. A}{55}{1997}{R1561}
\item{[6]}
\refer{R. Derka, V. Bu\v{z}ek, A. K. Ekert}{Phys. Rev. Lett.}{80}{1998}{1571}
\item{[7]}
K. Banaszek: {\sl ``Maximum-likelihood estimation of photon-number
distribution from homodyne statistics''}, to appear in {\sl Phys. Rev. A}.
\item{[8]}
\refer{U. Leonhardt, M. Munroe, T. Kiss, Th. Richter, M. G. Raymer}%
{Opt. Commun.}{127}{1996}{144}
\item{[9]}
\refer{Y. Vardi, D. Lee}{J. R. Statist. Soc.\ {\rm B}}{55}{1993}{569}

\end{description}

\end{document}